%% file: main.tex
\documentclass[letterpaper,twocolumn,10pt]{article}
\usepackage{usenix-2020-09}
\usepackage[small,compact]{titlesec}

\usepackage{tikz}
\usepackage{amsmath}
\usepackage{graphicx} 
\usepackage{framed}
\usepackage{array}
\usepackage[framemethod=TikZ]{mdframed} 
\usepackage{balance}
\usepackage{multirow, rotating, wasysym, url, multicol}
\usepackage[tight]{subfigure}
\usepackage[ruled,linesnumbered,vlined]{algorithm2e}
\usepackage{hyperref} 
\usepackage{color}
\usepackage[utf8]{inputenc}
\usepackage{tcolorbox}
\usepackage{caption}
\usepackage[english]{babel}
\usepackage{comment}
\usepackage[normalem]{ulem}
\usepackage{enumitem}
\usepackage{etoolbox}
\usepackage{tabularx}
\usepackage{float}

    \definecolor{zhao}{RGB}{205,133,0}

\usepackage[textsize=tiny,textwidth=0.6in]{todonotes}
\newcommand{\allnotes}[1]{}
\renewcommand{\allnotes}[1]{\textit{#1}}

\begin{document}

\date{}

\title{\Large \bf TSoR: TCP Socket over RDMA Container Network for Cloud Native Computing
}


\author{
{\rm Yulin Sun} \\
Quarksoft
\and
{\rm Qingming Qu} \\
Futurewei
\and
{\rm Chenxingyu Zhao} \\
University of Washington
\and
{\rm Arvind Krishnamurthy} \\
University of Washington
\and
{\rm Hong Chang} \\
Futurewei
\and
{\rm Ying Xiong} \\
Futurewei
}

\maketitle

\begin{abstract}
\input{zcxy/abstract.tex}
\end{abstract}

\input{introduction}
\input{background}

\input{zcxy/design.tex}

\input{implementation}
\input{evaluation}
\input{future_work}
\input{conclusion}




\newpage
\bibliographystyle{unsrt}
\bibliography{references}


\end{document}

%% file: zcxy/abstract.tex
 Cloud-native containerized applications constantly seek high-performance and easy-to-operate container network solutions. RDMA network is a potential enabler with higher throughput and lower latency than the standard TCP/IP network stack.  However, several challenges remain in equipping  containerized applications with RDMA network:
1) How to deliver transparent improvements without modifying  application code;
2) How to integrate RDMA-based network solutions with container orchestration systems;
3) How to efficiently utilize RDMA for container networks.

In this paper, we present an RDMA-based container network solution, TCP Socket over RDMA (TSoR), which addresses all the above challenges. 
To transparently accelerate applications using POSIX socket interfaces without modifications, we integrate TSoR with a container runtime that can intercept system calls for socket interfaces.
To be compatible with orchestration systems like Kubernetes, TSoR implements a container network following  the Kubernetes network model and satisfies all requirements of the model. To leverage RDMA benefits, TSoR designs a high-performance network stack that efficiently transfers TCP traffic using RDMA network. 
Thus, TSoR provides a turn-key solution for existing Kubernetes clusters to adopt the high-performance RDMA network with minimal effort. 
 Our evaluation results show that TSoR provides up to 2.3x higher throughput and 64\% lower latency for existing containerized applications, such as Redis key-value store and Node.js web server, with no code changes. TSoR code will be open-sourced. 
 

%% file: introduction.tex
\section{Introduction}

Container, the de facto computing unit for cloud-native applications, offers significant benefits in aspects of portability, efficiency, fault isolation, and ease of management\cite{containerization, mahgoub2022orion,sartakov2022cap,qiu2020firm,van2022blackbox, Groundhog}. Container plays an essential role in the evolving cloud computing paradigm. A wide range of applications has adopted containerization, including web servers\cite{nodejs}, databases\cite{redis}, machine learning \cite{zhou2019katib}, and more.    

A container orchestration system helps automate provisioning, deploying, networking, scaling, and managing containers. There are a few container orchestration systems among which Kubernets\cite{k8s} has become the de facto standard to deploy and operate containerized applications.

Container networking is a central part of any container orchestration system, which lays the foundation to deliver the benefits of containerization. Take Kubernetes as an example, there are multiple container networking solutions such as Flannel~\cite{Flannel}, Cilium~\cite{Cilium}, and Calico \cite{calico}. These solutions are based on the canonical TCP/IP protocol stack to provide overlay or underlay networks. 
As they add additional layers of virtualization over the kernel TCP/IP protocol stack,  their performance is bounded by the performance of the kernel TCP/IP network.

Remote Direct Memory Access (RDMA) \cite{rdma}, a kernel-bypass and hardware-offloading network technique, offers significantly lower latency and higher throughput than traditional TCP/IP based network. RDMA network has been widely deployed in cloud environments to support data-intensive applications such as storage, machine learning, and high-performance computing\cite{kong2022collie,zhang2022justitia,guo2016rdma}. 
It is appealing to build a container network solution over RDMA to improve container network performance.

However, there are three challenges of transferring application TCP traffic over RDMA in containerized environments:

1) \textbf{Transparency on accelerating applications with no code changes}. Currently, most containerized applications use standard POSIX sockets \cite{atlidakis2016posix} for communications, regardless of using synchronous manners such as HTTP/gRPC\cite{grpc} or asynchronous manners such as AMQP~\cite{amqparch, microservicecomm}. Given that RDMA Verbs (operations) and programming models are significantly different from the POSIX socket interface, the cost of rewriting existing applications using RDMA Verbs is high.  Furthermore, to benefit from the mature ecosystems of the POSIX socket, we believe most new applications will still choose the POSIX socket to do network programming. Thus, transparency for socket-based applications is essential to design RDMA-based container network solutions.

2) \textbf{Compatibility on integrating network solutions with orchestration systems}. Orchestration systems like Kubernetes bring in benefits by automating the deployment of containerized applications, which is a crucial driving force for the large-scale adoption of containers. Orchestration systems rely on container network solutions (plugins) to support network virtualization (e.g., namespace isolation), orchestration (e.g., IP assignment), and management (e.g., network policy enforcement). Thus, integrating new container network solutions with orchestration systems can significantly reduce the deployment effort. However, it is challenging for RDMA-based container networks to achieve compatibility with the network model of orchestration systems.  

3) \textbf{Performance on building data path with RDMA networks}. To address the above two challenges, it is inevitable to lay additional layers over RDMA. Although raw RDMA provides a high-throughput and low-latency network, it is challenging to build a high-performance network stack that fully utilizes the potential of RDMA.




Several solutions have been proposed to transfer TCP traffic over RDMA network, such as SMC-R~\cite{smcr} and SocksDirect~\cite{li2019socksdirect}. SMC-R modifies Linux Kernel to intercept TCP traffic. SocksDirect uses a user-space library to intercept TCP traffic. 
They both address the issue of transparency for socket-based applications. But neither of them addresses the compatibility with the Kubernetes container network model. 

For another line of work, they try to expose the RDMA interface to containers, such as FreeFlow\cite{kim2019freeflow}. FreeFlow is a virtual RDMA networking framework to expose native RDMA Verbs into containers. FreeFlow does not natively support the standard POSIX socket interface, which relies on RSocket\cite{rsocket} to transfer TCP traffic to RDMA with additional overheads. FreeFlow does not address the integration with orchestration systems like Kubernetes either.      

In this paper, we present TCP Socket over RDMA (TSoR), an RDMA-based container network solution that supports POSIX socket API and is compatible with the Kubernetes networking model.  
It has the following characteristics:

$\bullet$  \textbf{Support POSIX socket interface}: TSoR intercepts POSIX socket interfaces and transfers TCP traffic over the underlying RDMA network. Applications using the POSIX socket interface can run over TSoR without modifications while benefiting a high-performance RDMA network.

$\bullet$ \textbf{Integrate with Kubernetes}: TSoR implements Kubernetes Container Network Interface (CNI) to integrate with container orchestration systems. It meets the requirements of the Kubernetes container network model so that the existing Kubernetes cluster can adopt TSoR with minimal effort.

$\bullet$ \textbf{Operate high-performance RDMA-based data plane}: TSoR data plane transmits containerized applications' TCP traffic over underlying RDMA network. We carefully design TSoR's data plane to fully utilize RDMA network's high throughput and low latency capability.

We demonstrate TSoR's capabilities by integrating TSoR with one secure container runtime,  Quark~\cite{Quark}. We implement TSoR with about 30K lines of Rust code. We evaluate TSoR performance with network micro-benchmarks such as iPerf3~\cite{iperf} and popular services such as Redis~\cite{redis} and Node.js~\cite{nodejs}. Our evaluation results show that TSoR provides up to 2.3x higher throughput and 64\% lower latency for Redis compared with runC\cite{runc} using host network which is the performance upper bound for several combinations of runtime and TCP/IP based container network.



%% file: background.tex
\section{Background and Related Work}

\subsection{Container Network and Kubernetes CNI}
 Containerized applications communicate with each other using container network, a central part of container orchestrating system. Container orchestration systems like Kubernetes supports Container Network Interface(CNI) plugin\cite{k8scni}, for solutions providing network connectivity. There are several CNI implementations (plugins) for the Kubernetes cluster, such as Flannel \cite{Flannel}, Cilium \cite{Cilium}, Calico \cite{calico}, and Weave\cite{cnilist}.

 Currently, the container network solutions mentioned above build the data path based on the TCP/IP stack. It is well-known that the Linux kernel's TCP/IP stack have lots of overhead which is hard to achieve high throughput and low latency \cite{zhang2021demikernel,zhang2022justitia,marty2019snap,guo2016rdma,dalton2018andromeda, zhuo2019slim, ma2022survey}. Thus, existing container network solutions suffer from the bottleneck of the TCP/IP stack.

\subsection{RDMA network}
Remote Direct Memory Access (RDMA) allows one application to directly access another application's memory on a remote machine without interrupting either side's CPU. RDMA can achieve low latency and high throughput with a kernel bypass and hardware-offloading stack. The RDMA network has been widely deployed in cluster infrastructures to support data-intensive applications such as storage, machine learning, and High-Performance Computing (HPC) \cite{wei2018deconstructing,wei2020fast, burke2021prism,guo2016rdma, fork}. 

RDMA library provides API called Verbs for developers to write network programming code using RDMA network. Before performing RDMA operations, some steps including memory registration and connection establishment between nodes need to be done first. RDMA provides a concept called Queue Pair(QP) which is roughly equivalent to a TCP socket to accomplish this setup. Verbs API can be used to perform RDMA operations after QP is established.

 \begin{figure}[tbp]
     \centering
 \includegraphics[width=\linewidth]{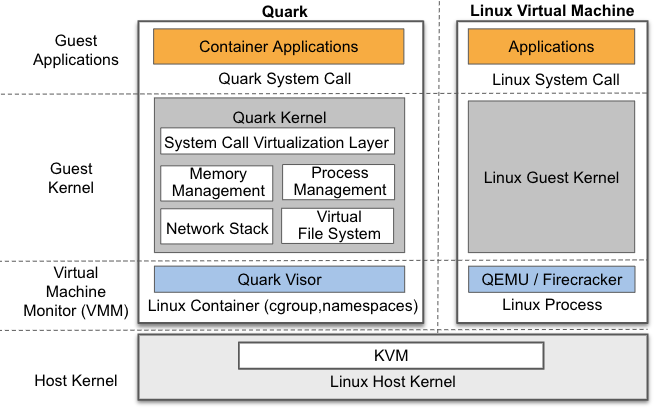}
 \caption{Quark secure container model  and LinuxVM.  \label{fig::quark-arch}}
 \end{figure}

 \begin{figure*}[ht!]
     \centering
 \includegraphics[width=0.9\linewidth]{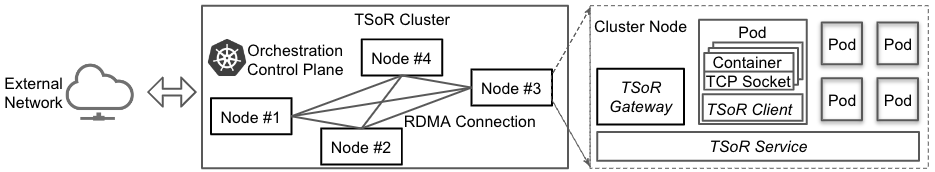}
 \caption{Architecture of TSoR.}
 \label{fig::arch}
 \end{figure*}

\subsection{Secure Container Runtime and Quark Architecture}

Secure container runtime performs like a typical container runtime\cite{containerruntime} with high efficiency and portability. Further, it provides isolation by building atop a lightweight virtual machine,  which is stronger than traditional containers' process-level isolation.  There is several  secure container runtimes such as Kata~\cite{kata, RunD}, AWS FireCracker~\cite{firecracker}, Google gVsior~\cite{gVisor}, and Alibaba runD \cite{RunD}.


Different secure container runtimes use different mechanisms to achieve secure isolation. Kata and Firecracker use Linux-kernel-based virtual machines. Quark and gVisor are secure container runtimes for serverless computing with a dedicated OS kernel and a Virtual Machine Monitor (VMM). They intercept a container application's system calls and provide a Linux-compatible system call interface. They support Container Runtime Interface (CRI)~\cite{cri} and  Open Container Initiative (OCI)~\cite{oci} so that existing Linux container images can run inside them without changes required. 

Figure~\ref{fig::quark-arch} shows Quark's architecture together with a Linux virtual machine. They both run over the Linux host kernel with the KVM hypervisor\cite{kvm}. The Quark runtime process runs in a standard Linux container with a Cgroup, network/file system namespace for isolation. Quark develops a new OS Kernel (QKernel) and VMM (QVisor) optimized for cloud-native applications. Quark implements most Linux kernel functions, such as network protocol stack, memory management, and process management. Quark is designed as a guest OS, so it relies on the host OS to manage physical devices, which saves the cost of complex IO and physical memory management. Compared with Linux VM-based container runtime, Quark has less memory footprint and lower startup latency. 

The Quark runtime intercepts containerized application's POSIX TCP socket API and implement them through host OS TCP/IP stack. We demonstrate TSoR capability by replacing Quark's original TCP protocol stack with RDMA stack.

%% file: zcxy/design.tex
\section{Overview}

\subsection{Design Goals}

We require our container network solution to satisfy the following design goals:

 $\bullet$ \textbf{High Performance}. We aim to provide a high-throughput and low-latency network for containerized applications, which can improve application performance.  

  $\bullet$  \textbf{Transparency to application}: For applications using standard POSIX socket interface, we aim to improve its performance without any code changes. 

  $\bullet$ \textbf{Compatibility with orchestration systems}. Our solution is designed to be fully compatible with existing orchestration systems like Kubernetes. We aim to meet all requirements of the Kubernetes networking model so that we can seamlessly and effortlessly integrate our solution with existing Kubernetes clusters and benefit from various toolkits from the Kubernetes ecosystem.

\subsection{Principle}
We highlight two principles to guide our design:

 \textbf{Separation of Concerns}: Although we aim at providing TCP socket over RDMA (TSoR), we do not seek the solution to tightly couple RDMA operation with container runtime. Instead, we build the TSoR module separately and then serve the TCP socket as a service(we call it as TSoR Service). There are several key advantages: 1) It can enable multi-socket sharing over one RDMA connection, which mitigates the scalability issues of RDMA. 2) It can enable the flexibility of enforcing the QoS policies for multiple sockets in one place. 3) It opens the opportunity for independent development of container runtime and TSoR service.

 \textbf{Don't Repeat Yourself}: We design TSoR following the orchestrating system networking model rather than the self-design model. Integration with the existing orchestrating system can standardize and simplify the design of the control plane. Further, our design can benefit from the formal and well-defined abstractions of the networking model from the orchestrating system. We demonstrate TSoR's capability by integrating with Kubernetes. 



\subsection{TSoR Architecture}




The architecture of TSoR is shown in Figure~\ref{fig::arch}, which depicts a cluster with four nodes using TSoR (We call it a TSoR Cluster, to be brief.). Cluster nodes connect to each other through an RDMA connection. External traffic,  by default using TCP, access the cluster via a gateway. Each cluster has a control plane from the orchestration system (Kubernetes is used as an example here). Each cluster node contains three TSoR components: TSoR Service, TSoR Client, and TSoR Gateway.

 \textbf{TSoR Service }: TSoR Service is TSoR's core component that serves the Pods\footnote{A Pod is a group of one or more containers with shared resources such as network namespace and IP address.} and TSoR gateways which are located in the same node as the TSoR Service. TSoR Service manages RDMA connections and executes data transmission to other nodes via RDMA NIC. TSoR Service also talks to the orchestration control plane, e.g., Kubernetes API Server.
 
\textbf{TSoR Client }: Each Pod or TSoR gateway has a TSoR client. After container runtime (We use Quark in this paper as an example) intercepts containerized applications' POSIX socket API calls, TSoR client will take over and talk to TSoR Service to set up the connection and transmit data over RDMA.

\textbf{TSoR Gateway }: TSoR cluster's internal traffic is transmitted over the RDMA network, while external traffic uses TCP/IP protocol. To connect Pods in the TSoR cluster with external resources, TSoR gateway performs protocol transformation between TCP and RDMA traffic. TSoR Gateway also uses TSoR Client to interact with TSoR Service.


Next, we will give a detailed description of TSoR Service (Section \ref{sec::tsor-serivce}), TSoR Client (Section \ref{sec::tsor-client}), TSoR Operation (Section \ref{sec::tsor-operation}), and TSoR Gateway (Section \ref{sec::tsor-gateway}).

\section{Design}
\subsection{TSoR Service}
\label{sec::tsor-serivce}







\begin{figure}[ht!]
     \centering
 \includegraphics[width=0.8\linewidth]{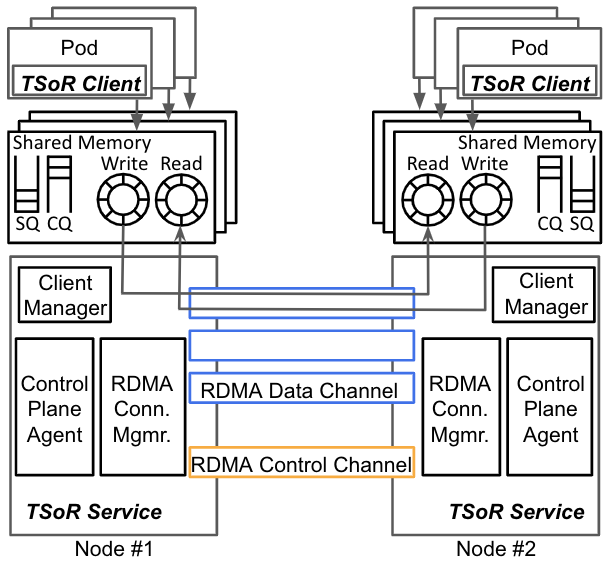}
 \caption{TSoR Service}
 \label{fig::rdma-service}
 \end{figure}
Figure \ref{fig::rdma-service} presents TSoR's core component, \textit{TSoR Service}, which consists of the components as follows. 

\subsubsection{Client Manager}

The Client Manager manages two types of TSoR clients: Pods and TSoR gateway. Client Manager uses shared memory to communicate with TSoR clients. The shared memory region maintains metadata and data structures associated with each TSoR client, which is described in TSoR Client section \ref{sec::tsor-client}.


\subsubsection{Control Plane Agent} TSoR Service needs to get connection related metadata from the orchestration system control plane, e.g. Kubernetes API Server in the Kubernetes cluster. The connection-related metadata includes active cluster node list, Pod list and cluster connection permission control policy. Based on the metadata, TSoR Service determines whether it is permitted to and how it sets up virtual TCP connections (which we refer to as an RDMA Data Channel) to map to real TCP connections in TSoR Clients.

\subsubsection{RDMA Connection Manager} 
\label{sec::rdma-cm}
RDMA Connection Manager handles RDMA Queue Pair (QP) connections which use Reliable Connection (RC) transport modes\cite{rdmaprogamming}. RDMA Connection Manager is responsible for creating and cleaning up QP connections. 
QP connection is used to transfer TCP traffic over RDMA. There are two main challenges for RDMA Connection Manager design:

\textbf{Challenge \#1: RDMA connection scalability}. In a typical cluster environment, micro-services within containers could set up a large number of concurrent TCP connections. However, the existing RDMA network falls short on the scalability issue, which is widely reported in \cite{kalia2016design, kong2022collie, wang2019vsocket, zhang2022justitia}. When the number of Queue Pair connections is large, the aggregated performance degrades dramatically. The root cause is the contention on the RDMA NIC's internal hardware resource, which is beyond the control of the container network. Given the high concurrency of scenarios using TCP connections, it is not practicable to build a one-to-one mapping between the TCP connection and the RDMA QP connection. 

\textbf{Challenge \#2: RDMA connection setup latency}: RDMA connection setup consists of creating resources (e.g. Queue Pair), exchanging metadata information and changing state\cite{rdmaprogamming}. Usually RDMA uses TCP connection as a communication channel to exchange Queue Pair metadata. We measured the latency of creating QP Connection based on TCP and found the latency is up to several milliseconds, while a typical TCP network only takes about hundreds of microseconds to establish a TCP connection. For scenarios with frequently establishing short-lived TCP socket connections, the time cost is significant if each short-lived TCP connection requires creating a separate RDMA connection.

We provide two solutions to tackle the aforementioned challenges:

\textbf{Solution \#1: Node-level RDMA connection multiplex}.
To solve the RDMA connection scalability issue, we multiplex the single long-lived RDMA connection for all TCP connections between the same pair of nodes.  Thus, the number of RDMA connections is determined by the number of cluster nodes which is orders of magnitude less than the number of TCP connections. RDMA connection multiplexing enables TSoR to support a large number of concurrent TCP connections.


 As Figure \ref{fig::rdma-service} shows, TSoR introduces a concept of RDMA Channel, which is mapped to a TCP socket connection. Logically the relationship between an RDMA Connection and an RDMA Channel is like the relation between a cable and wires inside the cable. Each end of RDMA channel has two data ring buffers: a read buffer and a write buffer.   
  The write buffer on one end connects to the read buffer on the other end. RDMA needs a virtually continuous memory block that is registered beforehand to work with. That memory block is called Memory Region(MR). The data ring buffers act as MR of RDMA Connection so that RDMA NIC can transfer data from the local write buffer to the remote peer's read buffer. Besides RDMA Data channels for data transfer, each pair of nodes maintains one RDMA Control Channel to exchange control messages. For example, nodes use control messages to tell remote peers about the available space of read buffer for RDMA data channels. Then remote peers can rate limit the data transfer according to the buffer space available.

\textbf{Solution \#2: Pre-established RDMA Connection}. Due to the long latency of RDMA connection setup, TSoR pre-establish RDMA connections rather than doing that when TCP connections are requested by user applications. When a node joins the cluster, TSoR Service on the node will start up and establish RDMA Connections to all its peer nodes in the cluster. At the time applications initiate TCP connections, the RDMA data channel can directly use the pre-established RDMA connection. By using the pre-established RDMA connection, TCP connection establishment does not need to pay the time cost for RDMA QP setup. Also, the low-latency and reliable data path of RDMA can speed up the handshake process of TCP. In Section \ref{sec::handshake}, we will describe the detailed process of TCP connection establishment.

TSoR service uses the following steps to establish an RDMA connection between cluster nodes:

$\bullet$ \textbf{Step \#1}: each RDMA Connection Manager starts a TCP server to listen on a known port to wait for other nodes' TSoR Service to connect; 

$\bullet$ \textbf{Step \#2}: When a new node joins the cluster, it retrieves an active cluster node list from the orchestration control plane;

$\bullet$ \textbf{Step \#3}: The newly joined cluster node initiates TCP connection with TSoR Service on existing peer nodes; 

$\bullet$ \textbf{Step \#4}: New node setup RDMA connections with existing peer nodes by exchanging QP metadata over the TCP connection.

\subsection{TSoR Client}
\label{sec::tsor-client}
\subsubsection{POSIX Socket API and System Call}
Containerized applications in TSoR Cluster can continue using POSIX socket APIs, which are fully supported by Quark container runtime. Quark runtime implements all system calls required by POSIX socket APIs, including system calls related to the virtual file system layer (e.g., $SysRead$) and TCP/IP stack (e.g., $SysConnect$).


There are two types of POSIX socket APIs: 1) Control primitives such as \textit{connect}, \textit{listen}, \textit{accept}, and \textit{close} are to set up/tear down the
TCP connections; 2) Data primitives such as \textit{read}, \textit{write}, \textit{send}, and \textit{recv} are to process data transmission. These primitives of the socket interface invoke system calls within the kernel of Quark container runtime.

Quark runtime intercepts POSIX socket related system calls. Then TSoR Client within the Quark runtime takes over those system calls, does transformation, and asks TSoR Service to execute them with RDMA connection. This way applications use POSIX socket APIs does not need any code change to utilize RDMA network.

In TSoR Operation section \ref{sec::tsor-operation}, we will describe how POSIX socket API calls are fulfilled along the whole stack end to end.




\subsubsection{Shared Memory between TSoR Client and Service}
\label{sec::shm-comm}

TSoR Clients communicate with TSoR Service using Shared Memory. For each client, RDMA Client Manager creates one shared memory region, which consists of two key data structures: A pair of message queues and ring buffers described as follows:

\textbf{Message queue pair}: The message queue pair consists of two shared memory queues: Submission Queue (SQ) and Completion Queue (CQ) \footnote{Here, SQ and CQ are used for communication between TSoR Client and TSoR Service, which are different from Queue Pairs of RDMA connection.}. The queues are lock-free shared memory queues. TSoR Clients send request messages to TSoR Service through SQ and receive response messages from TSoR Service through CQ.

\textbf{Shared Buffers}: Shared Buffers consist of read/write data buffers to store data applications send/receive. Shared Buffers are structured as fixed-size ring buffers associated with head and tail pointers.
Head and tail pointers can indicate the memory address for the producer and consumer manipulating the ring buffer. Head and tail pointers also can tell the availability of the ring buffer data. Shared buffers play a similar role as the socket buffers of standard TCP/IP stack.

In Section \ref{sec::rdma-cm}, we described a memory region needs to be registered for RDMA connection. Share buffers between TSoR Client and TSoR Service are mapped to the same memory allocated for the Memory Region (MR) for RDMA connection. There are two benefits to mapping the Shared Buffers to MR: 1) It can save the cost of memory copy from the shared buffer to the MR of the RDMA connection. 2) It can save the memory footprint. RDMA requires MR to be pre-registered and pinned (i.e. no swapping). While establishing a node-to-node RDMA connection, memory resource is allocated and pinned for the MR of the RDMA connection. 
\subsection{TSoR Operation}
\label{sec::tsor-operation}

\begin{figure}[tbp]
     \centering
 \includegraphics[width=\linewidth]{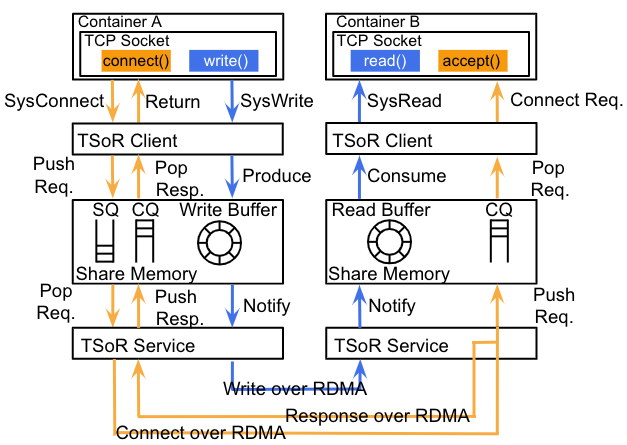}
 \caption{TCP connection setup and data transmission.}
 \label{fig::data-transmission}
 \end{figure}

\subsubsection{Data Trasmission}
Figure \ref{fig::data-transmission}  show the workflow of how two containers on different cluster nodes use TSoR to send/receive data which has the following main steps: 

$\bullet$ \textbf{Step \#1}: Containerized application calls $write$ which invokes the system call $SysWrite$. Container runtime executes the system call and interacts with the TSoR service as a TSoR client. TSoR Client plays as the producer for the write buffer by copying data from the application into the write buffer.

$\bullet$ \textbf{Step \#2}: TSoR clients enqueues one Write Request to the submission queue shared between the TSoR client and TSoR Service. Write Request is a signal to trigger the process of TSoR service. 

$\bullet$ \textbf{Step \#3}: TSoR service dequeues the Write Request and transmits data to the peer node over RDMA connection if the remote read buffer still has space. TSoR Service plays as a consumer for the local write buffer by transferring data to the remote peer's read buffer using RDMA IB verb.


$\bullet$ \textbf{Step \#4}: When TSoR Service is notified upon the arrival of data through RDMA completion event, it enqueues one request into the CQ between the TSoR Service and TSOR client to indicate the read buffer has new arrival data. 



$\bullet$ \textbf{Step \#5}: When the TSoR client is notified read buffer has data arrival, the TSoR client will consume data by copying data from the read buffer to the application buffer. The application finally receives the data from the remote peer.


We make a couple of optimizations for the above data transmission workflow:

$\bullet$ \textbf{Optimization \#1 Pipeling}: We enable pipelining for above Producer-Consumer workflow. While executing data transmission, on the local sender side, the TSoR client works as a producer in filling up the write buffer, and the TSoR service works as a consumer in reading the write buffer and then writes to the receiver-side read buffer using RDMA connection; on the remote receiver side, the TSoR service plays the role of producer for the read buffer, and the TSoR client consumes the data from the read buffer. On both the sender and receiver sides, the producer and consumer can work pipelining while transferring a stream of data packets. Such pipelining execution can increase the throughput of data transfer which is crucial to fully utilize the high-throughput RDMA connection.    

$\bullet$ \textbf{Optimization \#2 Signal Coalescing}: For the aforementioned transmission workflow of the general case, TSoR clients enqueue requests in Step-2 and TSoR service dequeues requests in Step-3. Such enqueue/dequeue operations incur the cost of pushing and polling the shared queues. We enable one mechanism called \textit{Signal Coalescing}: when TSoR clients detect the ring buffer is not empty as the application writes more data into the write buffer, TSoR clients do not enqueue the Write Request to notify TSoR service to process. To guarantee the data is sent out without the Write Request, TSoR Service will check if the ring buffer is empty after finishing sending data through the RDMA connection. If TSoR Service finds there is still data left in the ring buffer, TSoR Service continually sends data using the RDMA network even though without Write Request as a trigger signal. Via the \textit{Signal Coalescing}, TSoR client and service can save the cost of manipulating the shared SQ for most cases. SQ is only used to initiate the data transfer.

$\bullet$ \textbf{Optimization \#3 Idle Sleep and notification bitmap}: We enable the mechanism of idle sleep for TSoR Service. In Step-3, the TSoR service doesn't keep polling the submission queue to check whether there is a request from TSoR clients. During the idle time when there is no data to transfer, busy polling incurs the waste of CPU resources. To reduce such waste, the TSoR service uses a hybrid mode of busy pooling and event notification. When there is no data to transfer, the TSoR service sleeps until TSoR clients enqueue a request and then wake up the TSoR service via event notification. Once the TSoR service is woken up, it will do busy polling for a certain time before it sleeps again. TSoR service needs to support multiple TSoR clients, each with an SQ. It is wasteful to check every SQ whenever it is woken up. We implement a 2-layer bitmap for the TSoR service to do a quick look-up in order to find which TSoR client has requested for even further performance optimization. 

$\bullet$ \textbf{Optimization \#4 Lazy notification for read buffer available space}: In Step-3, TSoR service needs to check whether the remote read buffer has available space before sending data over RDMA. TSoR Service will keep track of the remote read buffer available space and stop sending data if the space is zero. When the application consumes data in the read buffer, the available space will increase, and TSoR service will notify the other side with the new space using the RDMA control channel. If the TSoR service sends a notification each time when an application reads a small amount of data, which is common for microservice communication, it is too chatty and can impact the overall performance. What we do is after the application reads data, TSoR client only requests TSoR Service to notify the other side's TSoR service when the read buffer's newly available space is more than half of the total ring buffer size.


\subsubsection{TCP Connection Establishment}
\label{sec::handshake}



We describe how TSoR clients establish TCP connections in this section. Typical TCP protocol uses a three-way handshake to establish a TCP connection which requires a Round-Trip Time (RTT)\cite{stevens2018unix}. As we mentioned in Section \ref{sec::rdma-cm}, TSoR can leverage the pre-established node-to-node RDMA connection to establish TCP connections.
Recall that we use the Reliable Connection transport mode for RDMA QP connection. We implement a two-way handshake for TCP connection establishment instead of three-way handshake like TCP. Also, RDMA network latency is lower than TCP/IP networks. So TSoR can benefit from the low-latency RTT of RDMA network to establish TCP connections.    

Figure \ref{fig::data-transmission} also shows the process of establishing a TCP connection under the setting of server-client where $Container~A$ is the client to initiate connection, and $Container~B$ is the server to accept the connection. A two-way handshake of establishing a connection works as follows:    

$\bullet$ \textbf{Step \#1:} $Container~A$ initiates a TCP connect request by calling $connect$ which invokes system call $SysConnect$. The Quark runtime intercepts $SysConnect$, obtains the destination IP address/port and let TSoR client do some tranformation and interact with TSoR Service via enqueuing one TCP connect request into SQ. 

$\bullet$ \textbf{Step \#2:} TSoR Service running on $Container~A$'s host pops out the requests from SQ and extracts the destination container IP and port. TSoR Service looks up the peer cluster node hosting $Container~B$(This look up 
 process will be explained in detail in Section \ref{sec::tsor-routing}), creates a local RDMA Data Channel and then sends a connection request using RDMA Control Channel over RDMA connection.  

$\bullet$ \textbf{Step \#3:} $Container~B$ waits on connection request via $accept$ socket API call. After TSoR Service running on $Container~B$'s host receives the connection requests, two tasks are executed to accept a connection request: Firstly, the TSoR service enqueues a Connect Request into the CQ to inform the $Container~B$ to establish a new connection.  Secondly, TSoR Service creates an RDMA Data channel and sends the response with read ring buffer information associated with the newly created RDMA Channel back to the peer node. 

$\bullet$ \textbf{Step \#4:} $Container~A$'s TSoR Service receives the response and then enqueue an TCP accept request into CQ to notify $Container~A$ that one connection is established and is ready for data transmission.

$\bullet$ \textbf{Step \#5:} TSoR Client in $Container~A$ pops out the response from CQ and then finishes the connection setup phase. The application is notified to be ready to send data.

 \subsection{TSoR Gateway}
 \label{sec::tsor-gateway}

\begin{figure}[tbp]
     \centering
 \includegraphics[width=\linewidth]{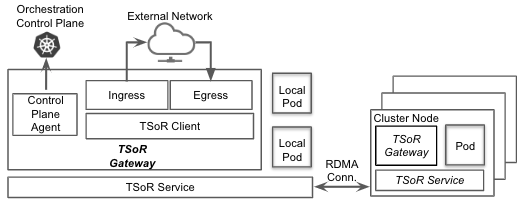}
 \caption{TSoR Gateway}
 \label{fig::tsor-gateway}
 \end{figure}

 TSoR gateway is an entry and exit point for TSoR cluster. Containers running in Pods of TSoR clusters need to access external services. Egress Gateway is an exit point through which Pod can connect to external resources. Also, the TSoR cluster needs to allow external resources to access services hosted in TSoR clusters. Ingress Gateway is an entry point for TSoR clusters. TSoR cluster's internal network traffic is over RDMA, while external traffic is over TCP/IP. So TSoR Gateway is to transform TCP traffic to RDMA Traffic  or vice versa.

 Figure \ref{fig::tsor-gateway} shows the TSoR Gateway architecture. The TSoR Gateway has the following main components:

$\bullet$ \textbf{TSoR Client}: Same as Pod, there is one TSoR Client running inside TSoR Gateway, which handles the communication with TSoR Service. 

$\bullet$ \textbf{Gateway Control Plane Agent}: TSoR Agent needs to get the network policy and connection metadata, such as mapping between external endpoint and internal Kubernetes service from cluster orchestration system control plane.

$\bullet$\textbf{Ingress TCP layer}: It exposes the internal service to external clients. It sets up a TCP server and listens to TCP ports that map to Kubernetes services. When it accepts an external TCP connection, it will create an RDMA data Channel and connect to one of the target internal Pods behind Kubernetes service using RDMA network.


$\bullet$ \textbf{Egress TCP layer}: It works as a proxy from an internal Pod to an external TCP-based service. It establishes a TCP connection to target the external TCP server for each internal RDMA channel, then transforms and forwards the internal RDMA traffic to external TCP traffic.
 
\subsection{Kubernetes Integration}




\subsubsection{Kubernetes Networking Model}

One design goal of TSoR is to provide a container network that is completely compatible with the Kubernetes Networking Model, which defines four main types of network communications:

$\bullet$ Pod-to-Pod Networking: In Kubernetes, all Pods can communicate with all other Pods as long as network policies (e.g., admission control rules) allow, regardless of whether the Pods are on the same cluster node or not.

$\bullet$ Pod-to-Service Networking: A service in Kubernetes cluster is an abstraction that exposes an application(a.k.a micro-service) running on a set of Pods with defined policies (e.g, load balancing) on how to access Pods. Service is assigned with an IP called Cluster IP.  The routing mechanism allows traffic addressed to the service Cluster IP to reach the actual Pods behind the service.

$\bullet$ External-to-Service Networking: Beyond TSoR clusters, we need to expose Kubernetes service to the external world like the Internet. Routing traffic to the TSoR cluster is complicated due to heterogeneity between TCP traffic outside of the cluster and RDMA traffic inside the cluster.

$\bullet$ Pod-to-External Networking: To allow Pods access resources and services (e.g., Internet) outside of the cluster, we need a mechanism to transform RDMA traffic inside the cluster to external TCP traffic.

Next, we present our routing solution to meet the above requirements and achieve compatibility with the Kubernetes networking model.

 

\subsubsection{Routing}
\label{sec::tsor-routing}
\begin{figure}[tbp]
     \centering
 \includegraphics[width=\linewidth]{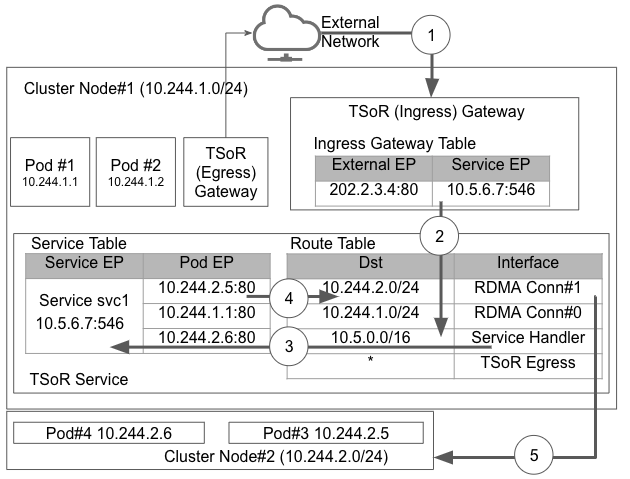}
 \caption{TSoR Container Network Routing. In this example,  TSoR assigns the subnet of 10.244.1.0/24 to Node1 while Node2 uses the subnet of 10.244.2.0/24. Pod1 running on Node1 gets IP 10.244.1.1, and Pod3 running on Node2 gets IP 10.244.2.5. The TSoR cluster exposes the service svc1 to the public with the external endpoint (EP) IP 202.2.3.4. The Ingress Gateway Table converts the external endpoint to the service endpoint 10.5.6.7:546. Route Table maps 10.5.6.7 to Service table. The Service handler load balances the service to the Pod with 10.244.2.5. Route Table maps 10.244.2.5 to locate the RDMA Connection \#1 to the Pod3 in the Node2.}
 \label{fig::routing-table}
 \end{figure}
Figure \ref{fig::routing-table} shows an example of the network configuration for TSoR Kubernetes cluster which depicts TSoR's routing solution to achieve compatibility with the networking model described before.

\textbf{Pod IP Assignment}: TSoR cluster assigns an IP address to each Pod. The TSoR control plane in the Kubernetes orchestration system will allocate and maintain the IP addresses.  While assigning IP addresses, TSoR uses a mechanism of \textit{static allocation}: TSoR allocates a fixed subnet CIDR (Classless Inter-Domain Routing) for each cluster node. 
Because each node has a fixed subnet CIDR, TSoR Service can maintain a relatively static mapping between the subnet CIDR and cluster node, which is only updated when nodes join or leave the TSoR cluster. Such static mapping is recorded by a Route Table described later. When TSoR Service gets a TCP connection request, it can simply look up the Route Table based on the subnet CIDR of the destination IP to get the destination node. Then, TSoR uses an RDMA connection associated with the destination node to establish a TCP connection. 
Static allocation can achieve low latency Pod startup by eliminating the Pod's IP address propagation. If the IP assignment of Pods takes a dynamic manner instead of the static allocation, the startup Pod has to wait for its IP assignment to be propagated across the cluster before serving user requests.
Low latency Pod startup is crucial for serverless computing, such as Function-as-a-Service (FaaS), which has no pre-provisioned resource and Pod is launched by user request invocation. IP address propagation resides in the critical path of end-to-end user request-response process and increases the latency.


Based on the IP address assigned to each Pod, we define three tables to route traffic for different types of communications, as Figure \ref{fig::routing-table} shows.  

$\bullet$ \textbf{ Ingress Gateway Table}: The table records the entries mapping the external facing IP and port to the internal service cluster IP and port. We call the tuple of IP and Port an Endpoint, abbreviated as EP, from now on. As defined by the Kubernetes networking model, the TSoR cluster can expose a Kubernetes service to the external world. External clients access the service via the external EP. By looking up the Ingress Gateway Table, TSoR converts the external EP to the internal EP.


$\bullet$ \textbf{ Cluster EP Table}: The table records the entries mapping the internal service EP to a set of Pod EP. The load balancing policy(e.g., round robin) defines which Pod will be selected to serve the request. After looking up the Service Table, the internal server EP is converted to a Pod EP.


$\bullet$  \textbf{Route Table}: The table records the entries mapping the subnet CIDR to different TSoR communication modules. There are three types of modules: RDMA connections, the Service handler, and the TSoR Egress Gateway. 



Based on the three tables, TSoR addresses different communications needs as follows:

\noindent \textbf{ Type-1 Pod to Pod}: This is the basic requirement of the container network. It includes intra-node communication, such as Pod\#1 to Pod\#2, and inter-Node communication, such as Pod\#1 to Pod\#3 in Figure \ref{fig::routing-table}. Based on TSoR Route Table, when the TSoR Service gets a connection request, it will find RDMA Connection for the destination based on the destination IP address, and then it can send the connection request to the destination. 

\noindent \textbf{Type-2 Pod to Kubernetes Service}: Just like Pod to Pod communication, TSoR Service will look up the TSoR Route Table at first. When it gets a service handler based on the destination endpoint, it will look up the Service Table and get a list of Pod IP endpoints. Then it will choose one of them based on the load balance policy as the destination endpoint and use the Pod to Pod communication path to send the connection request to it.

\noindent  \textbf{Type-3 External to Kubernetes Service}: TSoR supports the external to Kubernetes Service communication based on TSoR (Ingress) Gateway. When the TSoR (Ingress) Gateway gets a TCP connect request, it will look up the TSoR Ingress Gateway Table, and it will get the target Kubernetes Service endpoint. Then it will use the Pod to Kubernetes Service communication path to send the connection request.

\noindent \textbf{Type-4 Pod to external}:  Just like Pod to Pod communication, TSoR Service will forward the connection request to the TSoR (Egress) Gateway as a common Pod. When TSoR (Egress) Gateway gets a connection request, based on the target external IP address of the connection request, TSoR (Egress) Gateway will initiate a TCP connection to it. 

In addition to Kubernetes container network model requirement, TSoR also supports the following features:

\noindent \textbf{Network Policies and QoS Enforcement.} 
The network policy is a collection of rules which specify whether a Pod is allowed/blocked to access other Pods and services. Kubernetes  provides NetworkPolicy API  
to control traffic flow at the IP address or port level. TSoR Service is a convergence point of all traffic of the cluster Node.  TSoR Service enforces the network policy by adding network policy rules tables and checking them for each TCP connection establishment request.  TSoR supports QoS (Quality of Service) by implementing rate limiters in TSoR Service.

\noindent \textbf{Multi-tenant network secure  isolation}. TSoR supports multi-tenant network secure isolation, from which a Pod can not talk to another Pod of a different tenant. TSoR achieves this by assigning a tenant Id when assigning Pod IP. TSoR will block any TCP connection between different tenant\_id;

%% file: implementation.tex
\section{Implementation}

We implement the TSoR service and TSoR gateway from scratch while modifying the Quark runtime code to embed the TSoR client to communicate with the TSoR service.
Implementing a highly performant TSoR needs lots of considerations which are reflected in multiple places as follows:


$\bullet$ {\bf Programming language}: We choose Rust as the programming language instead of C/C++ or some managed language even though Rust has a steep learning curve. TSoR can't use managed languages as it needs access and manage memory directly and can't afford performance issues introduced by garbage collection. Also, Rust offers memory-safety and thread-safety guarantees with its own programming paradigm which is crucial for the reliability of a system-level component such as TSoR. One overhead to choosing Rust is we have to build our own rust library to export and wrap RDMA native library in C language while we still believe this choice can benefit TSoR in the long term. Total lines of Rust code for TSoR are around 30k to date. 

$\bullet$ {\bf Asynchronous programming everywhere}: RDMA can achieve single-digit microseconds latency end to end\cite{rdma}. Microseconds-level programming can't afford to block anywhere in the call stack. All functions of TSoR are event-driven in an asynchronous way. TSoR not only relies on TCP socket event to handle TCP-related operations\cite{stevens2018unix}, it also leverages RDMA's event to get a notification. e.g., when there's RDMA traffic arriving. To achieve this, we have to set up a TCP connection ourselves to exchange information necessary to establish an RDMA connection instead of using RDMA CM library\cite{rdmaprogamming} directly as RDMA CM  doesn't expose the file description we need to monitor RDMA event. 

$\bullet$ {\bf Lock-free data structures and synchronization between processes}: TSoR uses a lock-free queue pair (SQ and CQ) as a communication mechanism between TSoR client and service. TSoR client and service also use lock-free ring buffer to share data buffer memory. Both of them aim to eliminate contention between different processes. While it can achieve better performance, it also brings challenges. Because different CPU cores and memory layers (register, cache, and main memory) can be out of sync if loose memory ordering than necessary is chosen\cite{rustorder}. This kind of bug can be fairly difficult to reproduce and debug. We still choose this way considering its performance gains.

$\bullet$ {\bf RDMA operation selection}: RDMA supports different operations such as SEND, RECEIVE, and WRITE, etc \cite{rdmaprogamming}. We choose RDMA Write With Immediate (WRITE\_IMM). WRITE\_IMM operation could copy a user-space buffer data to a remote machine application buffer as a common WRITE operation together with a 32 bits immediate value. The immediate data can serve as a key for RDMA Channel to help TSoR multiplex TCP socket connections over one RDMA Connection and achieve lower latency. 







%% file: evaluation.tex
\section{Evaluation}






\begin{figure*}[ht]
	\centering
	\subfigure[Throughput]{
		\begin{minipage}[t]{0.32\textwidth}{
				\includegraphics[width=1\textwidth]{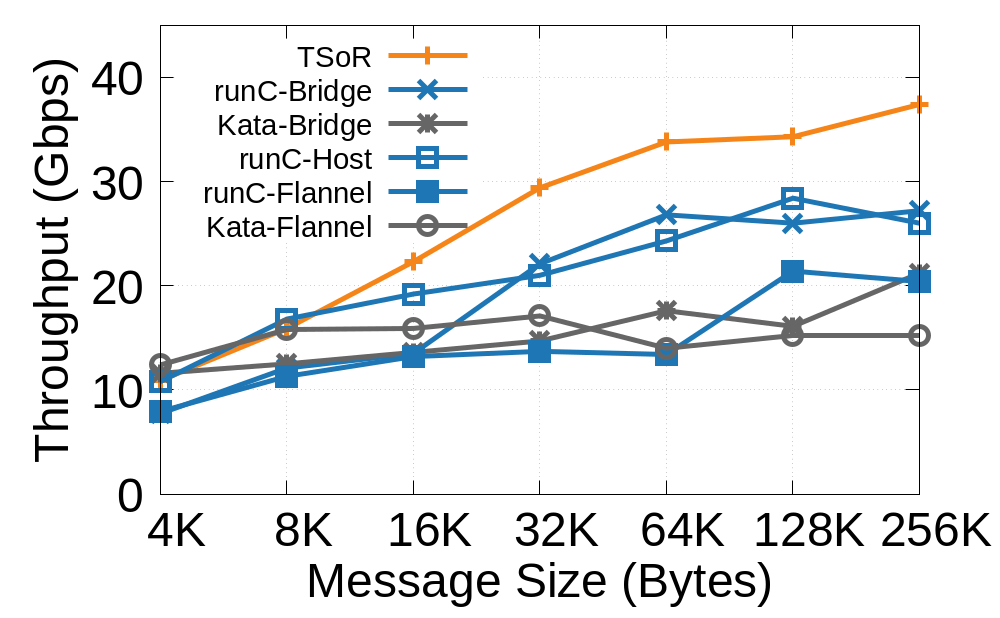}}
            \label{fig::single-throughput}
    	\end{minipage}}
        	\subfigure[Latency]{
		\begin{minipage}[t]{0.32\linewidth}{
				\includegraphics[width=\linewidth]{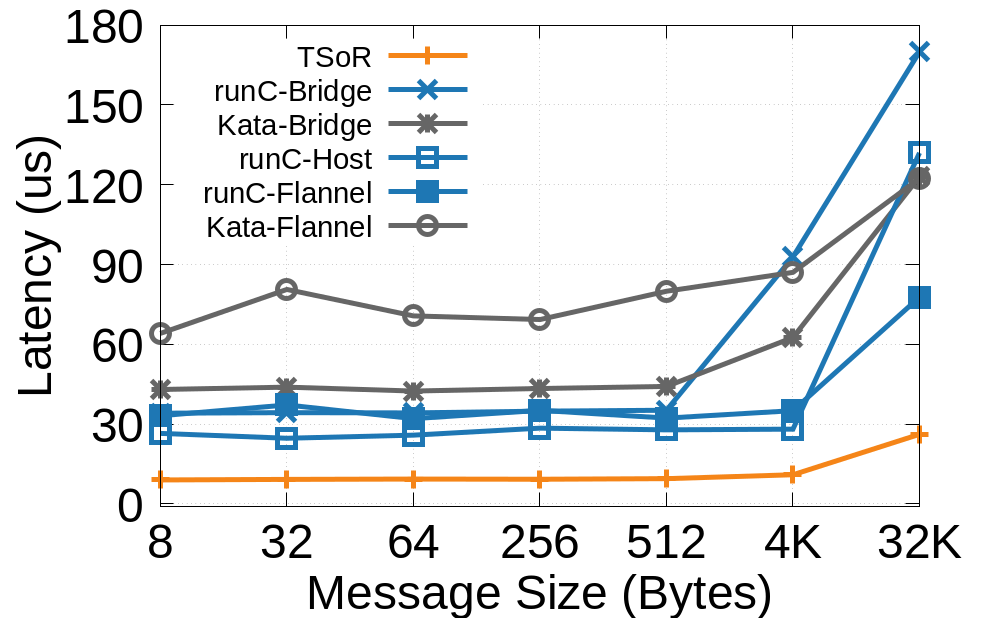}}
            \label{fig::single-latency}
    	\end{minipage}}
    	\subfigure[Multi-connection Throughput]{
		\begin{minipage}[t]{0.32\textwidth}{
				\includegraphics[width=1\textwidth]{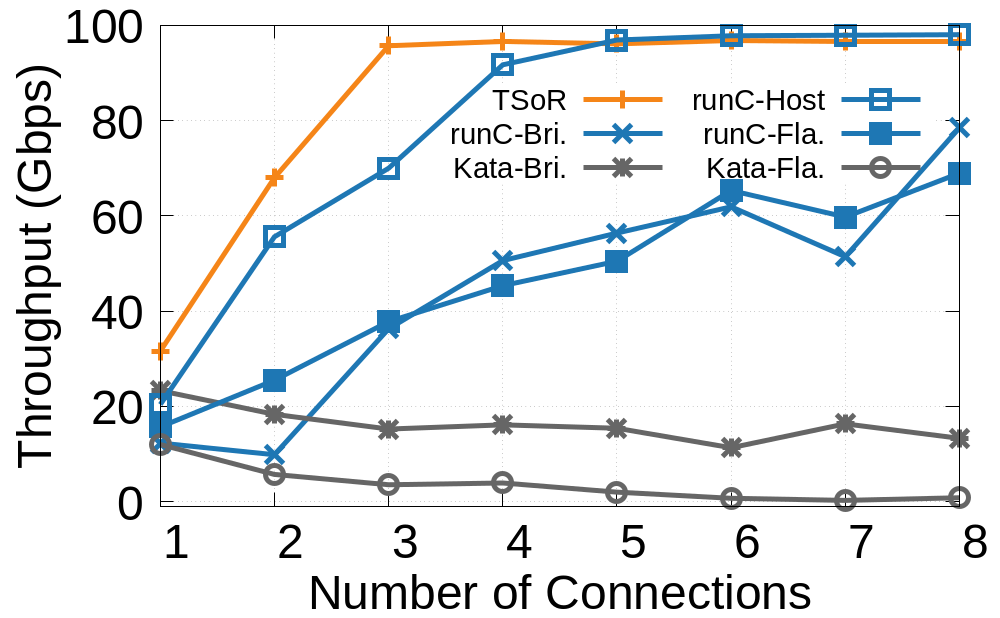}}
            \label{fig::multi-throughput}
    	\end{minipage}}
    \vspace{-5pt}
\caption{TCP connection throughput and latency between a pair of containers across hosts.}
\end{figure*}

\subsection{Setting}
\ 

\textbf{Testbed}. Our testbed comprises an RDMA-capable cluster with x86 servers connected to a 100Gbps Arista 716032-CQ switch. Each server has two Intel Xeon Gold 5218 processors, 96GB memory, and a 100Gbps dual-port Nvidia ConnectX-6 Dx NIC. We run Ubuntu 20.04 with kernel 5.15 and RoCEv2 for RDMA. We use Kubernetes v1.21.7 and Docker v20.10.21.

\textbf{Comparison Baselines}. We choose baselines based on the combination of two dimensions: container runtime and network mode. 

{\bf Container runtime}: 1) runC  \cite{runc} is a widely-used Open Container Initiative (OCI) runtime that can integrate with Docker and Kubernetes. runC is a host OS process with lightweight isolation, such as Cgroup and Linux namespaces. Its performance is better than VM-based secure container runtime such as Kata/Quark. runC can be configured to work with different container network solutions, including host network, bridge network, and Container Networking Interface (CNI) plugins like Flannel.  We use runC version 1.1.4 for experiments.  

2) Kata \cite{kata} is a container runtime that allocates containers inside a lightweight virtual machine using QEMU/KVM with its own lightweight Linux Kernel. Kata can be configured to work with bridge network and Flannel. As far as we know, Kata does not support using the host-mode network. We use Kata Container version 1.13 for experiments. 

{\bf Network mode}: 1) Host Network: In this mode, Containers share network namespace with the host OS. Containers share the TCP/IP stack within the host machine’s Linux kernel. Due to removing the network virtualization and security layers, the host network can provide better performance than other major container network solutions \cite{suo2018analysis}, yet lack isolation and security mechanisms. 

2) Bridge Network: Bridge is a network device in the data link layer which connects multiple network segments. Docker, by default, uses a bridge that allows containers to communicate with each other by connecting to the same bridge. We use Docker's default bridge for experiments. 

3) Flannel Network  \cite{Flannel}: Flannel is a container network solution designed for Kubernetes, which creates an overlay network over the host network. Each container has its own IP address. Flannel runs a daemon process that uses a kernel route table to achieve IP address translation while the container communicates across hosts. We use Flannel version 0.3.1 and set it up as CNI using Kubernetes for experiments.




\subsection{Microbenchmarks}
\textbf{Throughput}: Figure \ref{fig::single-throughput} presents a throughput comparison for a single TCP connection between a pair of containers across hosts. We use iPerf3 to generate TCP traffic while varying the message size. We focus on the throughput performance for transferring the message with a relatively large size (larger than $4KB$). Because in practice, most of the throughput-hungry scenarios like file transferring use relatively large message sizes. Varying message size from $4KB$ to $256KB$, TSoR achieves higher throughput compared with other combinations. As message size increases, TSoR's benefits are more significant. Because the RDMA-based data path of TSoR has a higher upper limit of throughput than others which are bounded by the host OS TCP/IP stack.
For a typical message size of  $64KB$, TSoR achieves 34.3 Gbps throughput, which is even higher than host-mode runC viewed as the performance upper bound for container network solutions. As we know, host-mode networks are not practical for real-world deployments due to the challenges of management and security.

\begin{figure}[bhp!]
	\centering
	\includegraphics[width=0.8\linewidth]{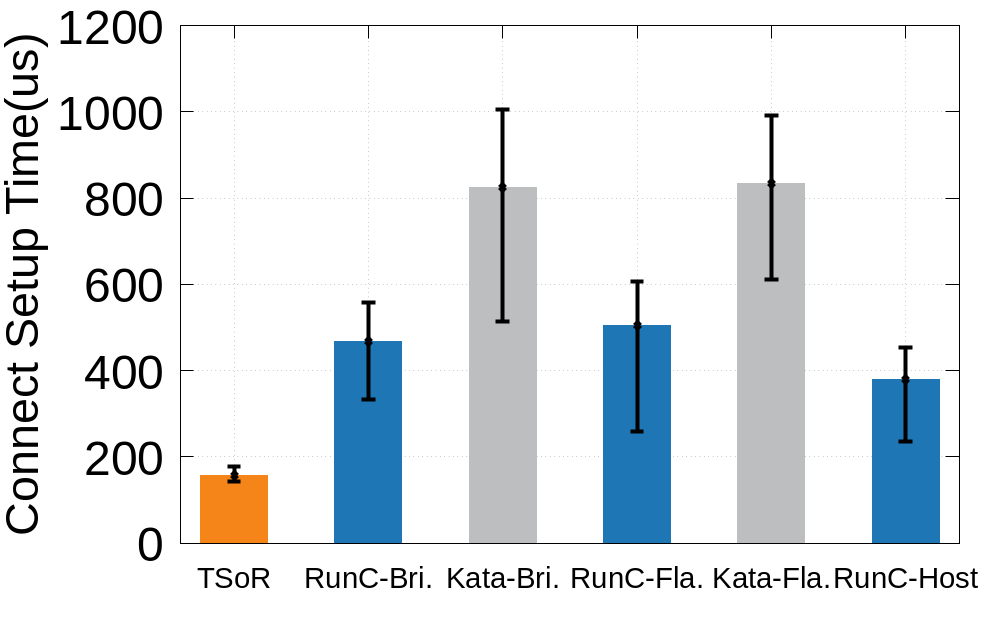}
 \vspace{-10pt}
\caption{TCP Connection Setup Time. Error bars show the min/max setup time.}
 \label{fig::conn-setup}
\end{figure}

\begin{figure*}[ht!]
	\centering
    \subfigure[Single Connection]{
		\begin{minipage}[t]{0.32\linewidth}{
				\includegraphics[width=\linewidth]{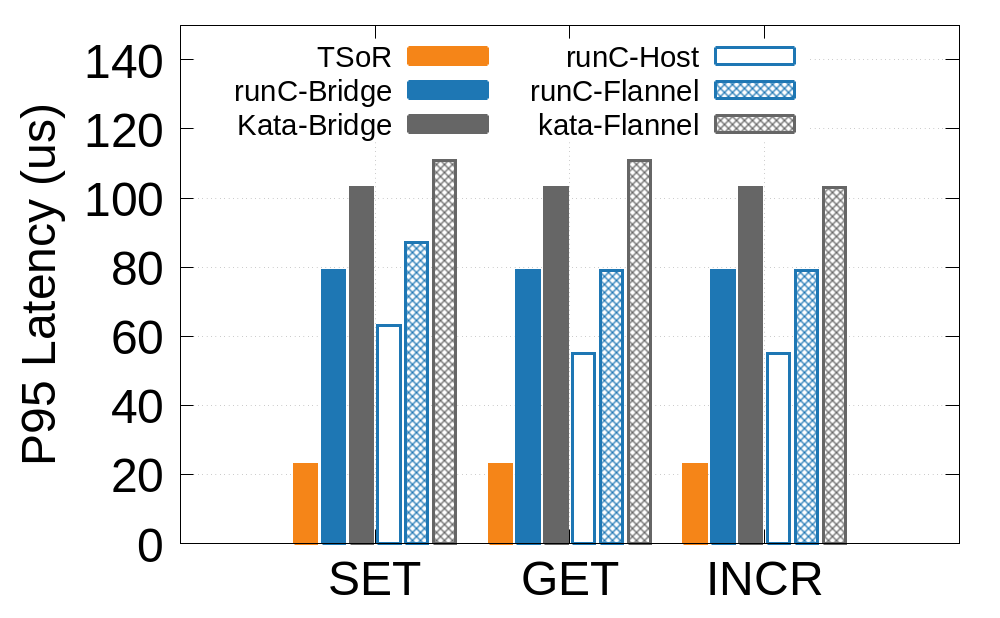}}
            \label{fig::redis_single_lat}
        \vspace{-10pt}
    	\end{minipage}}
    \subfigure[Multi-connection GET]{
		\begin{minipage}[t]{0.32\linewidth}{
				\includegraphics[width=\linewidth]{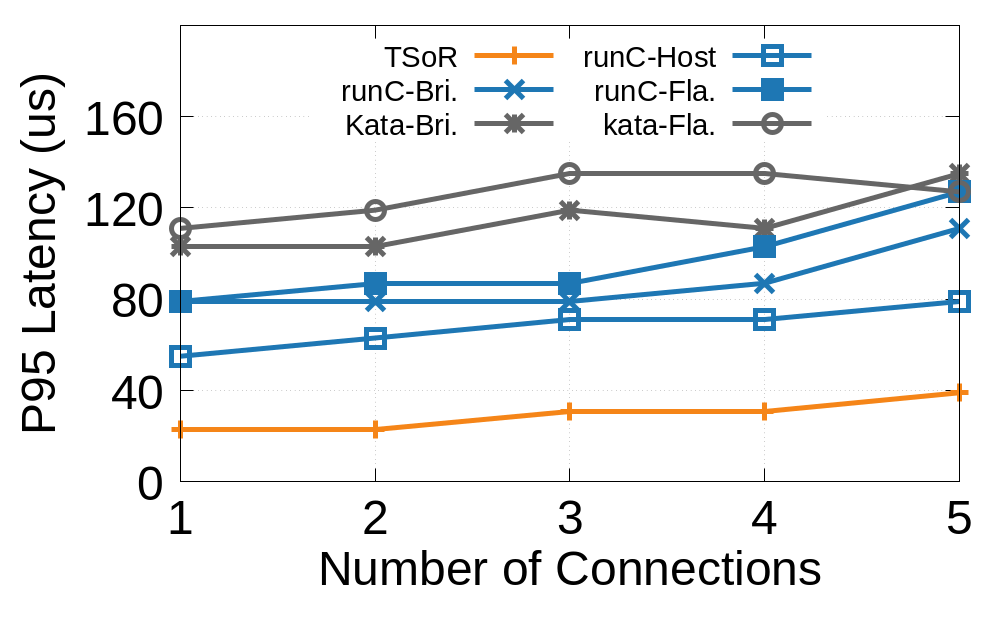}}
            \label{fig::redis_multi_get_lat}
                \vspace{-10pt}
    	\end{minipage}}
    \subfigure[Multi-connection SET]{
		\begin{minipage}[t]{0.32\linewidth}{
				\includegraphics[width=\linewidth]{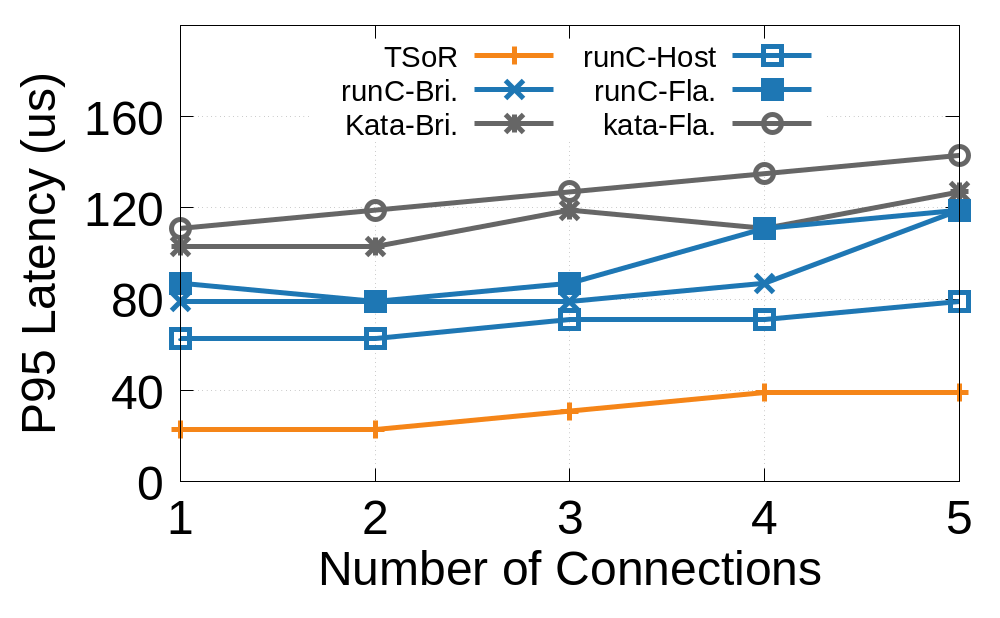}}
             \label{fig::redis_multi_set_lat}
            \vspace{-10pt}
    	\end{minipage}}
 
    \subfigure[Single Connection]{
		\begin{minipage}[t]{0.32\linewidth}{
				\includegraphics[width=\linewidth]{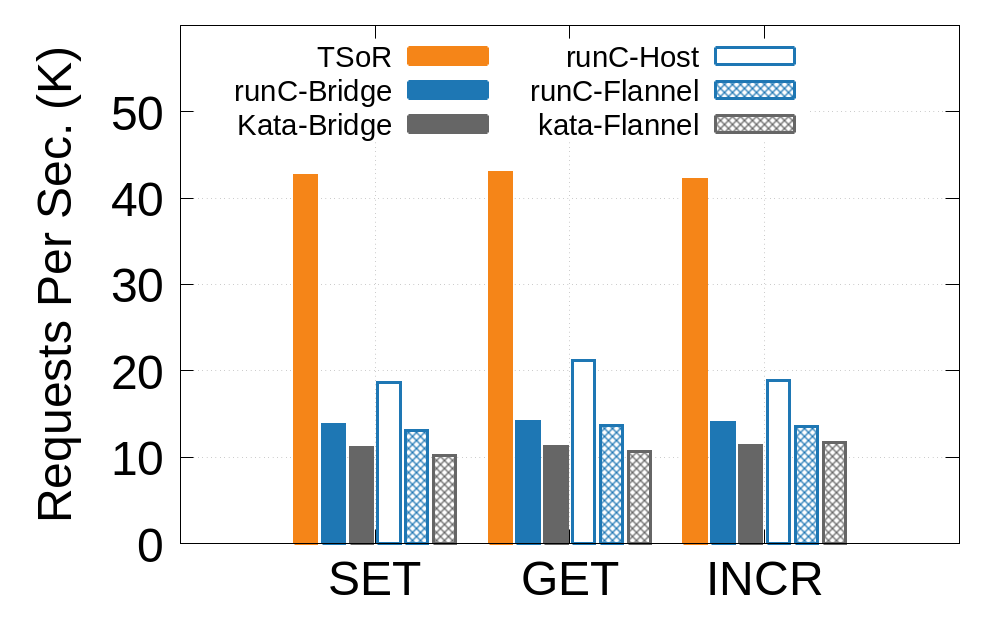}}
            \label{fig::redis_single_rps}
       \vspace{-10pt}
    	\end{minipage}}
            \subfigure[Multi-connection GET]{
		\begin{minipage}[t]{0.32\linewidth}{
				\includegraphics[width=\linewidth]{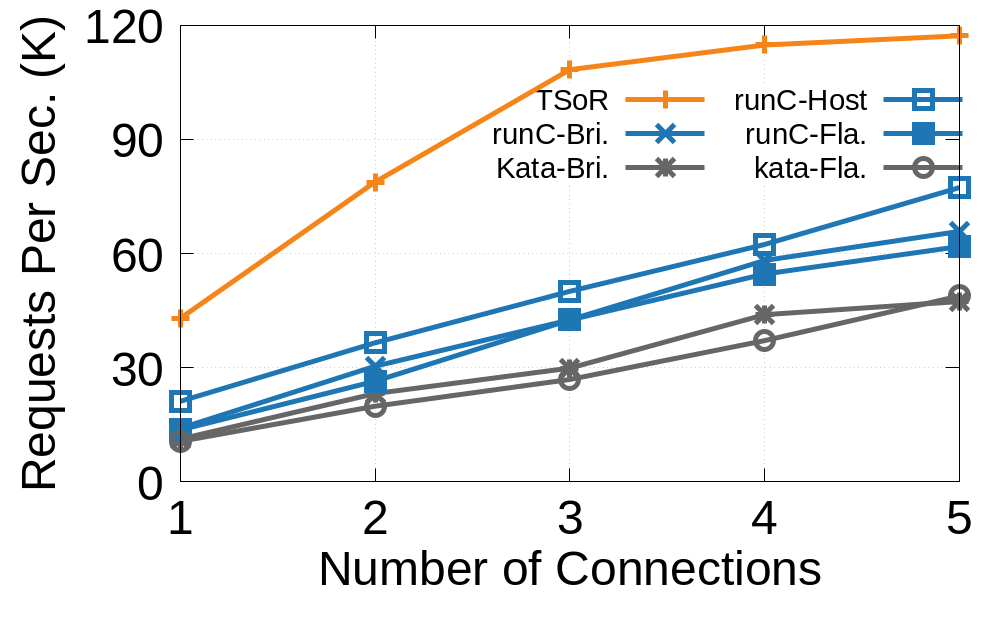}}
            \label{fig::redis_multi_get_rps}
         \vspace{-10pt}
    	\end{minipage}}
            \subfigure[Multi-connection SET]{
		\begin{minipage}[t]{0.32\linewidth}{
				\includegraphics[width=\linewidth]{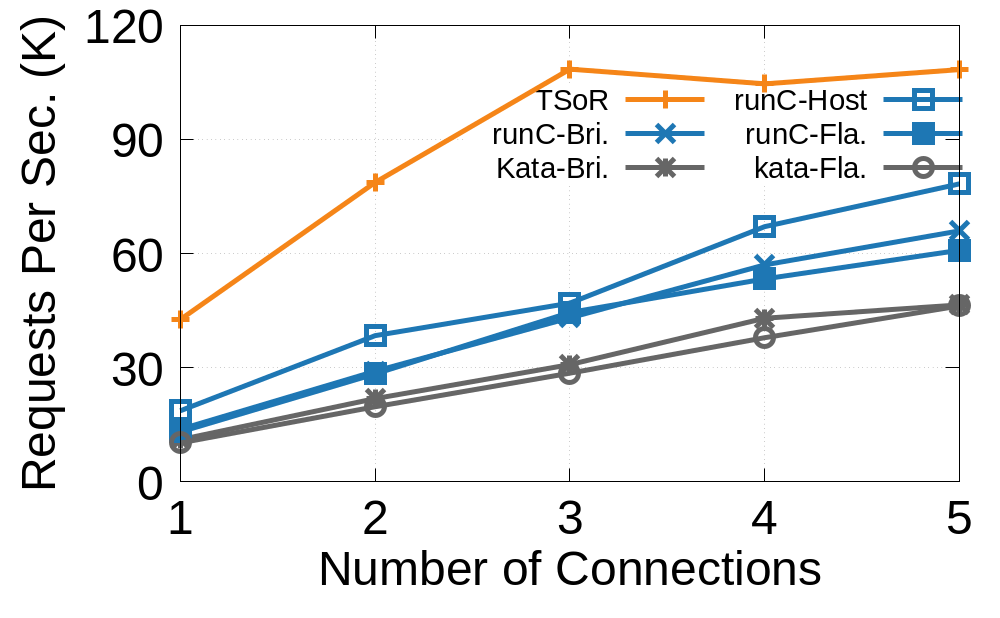}}
             \label{fig::redis_multi_set_rps}
            \vspace{-10pt}
    	\end{minipage}}
\caption{Redis Performance. TSoR achieves lower response latency, higher throughput, and  better scalability.}
\label{fig::redis}
\end{figure*}

\textbf{Latency}: Figure \ref{fig::single-latency} shows latency comparison for a cross-host TCP connection. We use NPtcp to measure the average latency for executing 1000 times message transferring with varying the message size. We focus on the latency performance for a small message which is common in typical microservices using RPCs. TSoR achieves constantly lower latency for small messages. For 64-byte messages, TSoR achieves 9.3 us latency which is 70\% lower than runC with a host network. Due to removing lots of layers of security and virtualization, runC with host network plays as a performance ceiling for host OS TCP/IP stack-based container network solutions. By building a data path atop RDMA, TSoR avoids the bottlenecks of the Linux kernel's TCP stack and thus outperforms runC with the host network.

\textbf{Multi-connection Scalability}: Figure \ref{fig::multi-throughput} presents a throughput comparison for 64-KB messages while scaling the number of connections from 1 to 8. On both the sender and receiver sides, we pin each connection thread on a core. We can see that TSoR achieves almost linear scalability and saturates 100Gbps with only three connections. runC with a host network can achieve 100Gbps with more connections. Due to internal resource contention and virtualization overhead within bridge-mode and flannel solutions, runC can not scale to the 100Gbps line rate. Due to the resource sharing within the Kata runtime, the multi-connection throughput of Kata is even lower than the single connection.

\textbf{TCP Connection Establishment Time}. Figure \ref{fig::conn-setup} shows the TCP connection establishment time. Using TSoR, the establishment of a TCP connection takes a much shorter time. TSoR only needs less than half of the establishment time compared with host-mode runC. As we described before, although TSoR requires one round trip between server and client, similar to the handshake process of standard TCP, RDMA has a much lower round trip latency. So TSoR can benefit from RDMA to establish connections faster. We notice that TSoR's establishment time has much less variation compared with others. Because RDMA's low latency performance is more stable than the host OS TCP/IP stack. As the experiment shows,  TSoR can more efficiently handle connection establishment for many short-lived TCP connections, which is challenging for lots of prior work \cite{zhuo2019slim,li2019socksdirect}.

\begin{figure*}[ht!]
	\centering
    \subfigure[Node.js Latency]{
		\begin{minipage}[t]{0.32\linewidth}{
				\includegraphics[width=\linewidth]{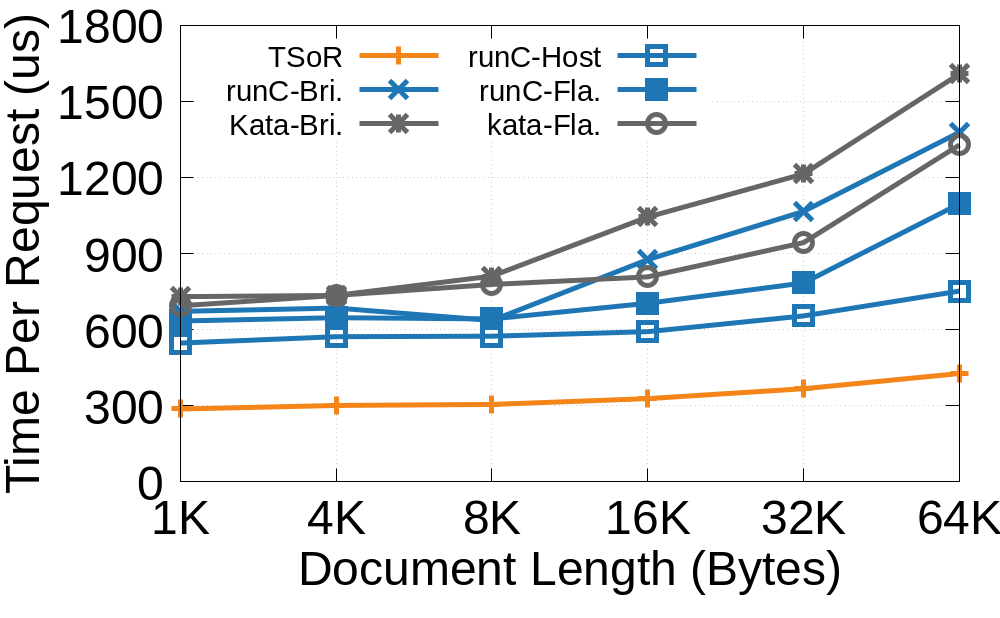}}
            \label{fig::nodejs_lat}
             \vspace{-10pt}
    	\end{minipage}}
    \subfigure[Node.js Throughput]{
		\begin{minipage}[t]{0.32\linewidth}{
				\includegraphics[width=\linewidth]{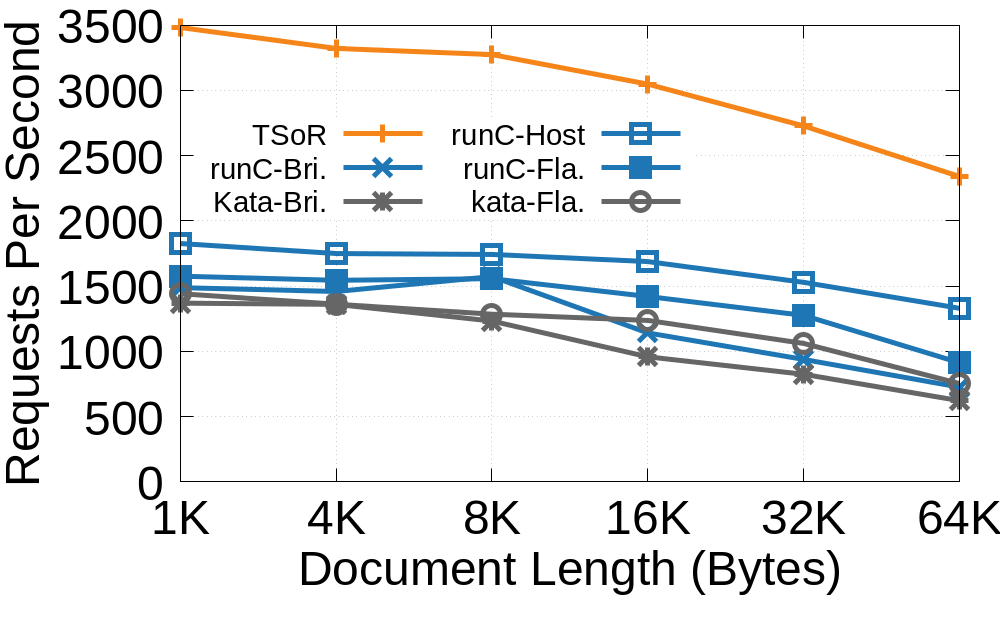}}
     \label{fig::nodejs_rps}
      \vspace{-10pt}
    	\end{minipage}}
    \subfigure[Etcd Throughput]{
		\begin{minipage}[t]{0.32\linewidth}{
				\includegraphics[width=\linewidth]{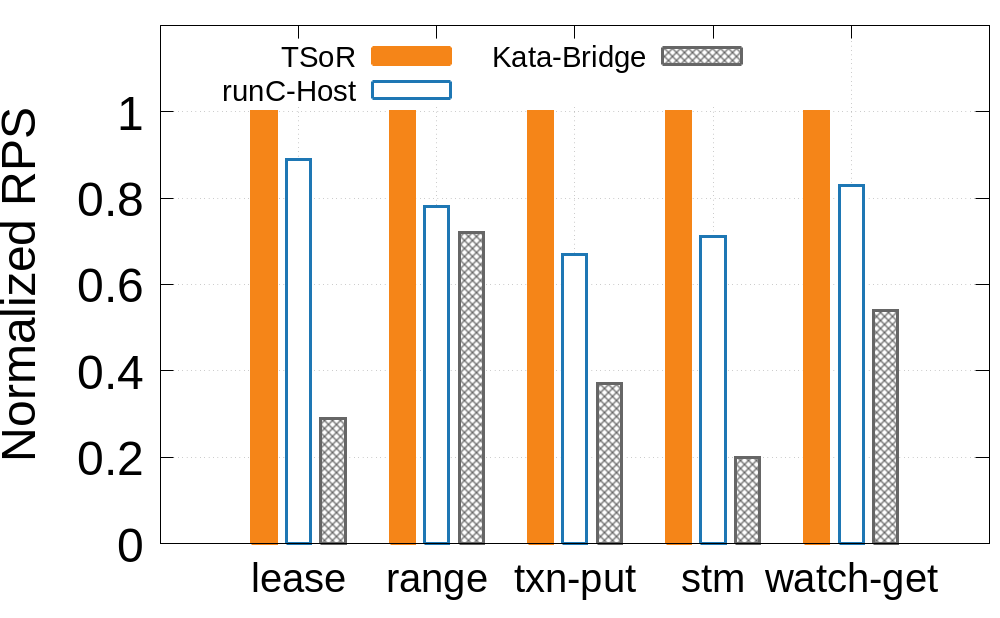}}
     \label{fig::etcd}
      \vspace{-10pt}
    	\end{minipage}}
   \vspace{-10pt}
\caption{Node.js and etcd performance. TSoR achieves lower response latency and higher throughput.}
\end{figure*}

\subsection{Real-world Applications}
\subsubsection{Redis}
Redis is an in-memory key-value data store. We use the official Redis docker image (v7.0.5) and the built-in Redis-benchmark utility to collect the metrics of latency and throughput while running with different combinations.

Figure \ref{fig::redis} shows the response latency and throughput collected from the Redis benchmark. We set up the server and client in different containers deployed on different hosts. We select SET, GET, and INCR operations where INCR means incrementing the number stored at \textit{key} by one. 

Figure \ref{fig::redis_single_lat} presents the latency for executing SET, GET, and INCR operations with the default data size of 3 bytes.
To focus on measuring network performance and avoid internal queuing within the application, we set up a single connection for this experiment. For SET, TSoR's 95\% latency is only 23 us, 64\% percent lower than host-mode runC. Bridge-mode runC and Flannel-mode runC have a longer latency  than host-mode runC. And Kata's latency is longer than runC. These observations are consistent with some prior measurements \cite{cochak2021runc, kumar2020performance}. For all test operations, TSoR achieves significantly lower latency than others. 

Figure \ref{fig::redis_multi_get_lat} and Figure \ref{fig::redis_multi_set_lat} show the latency while increasing the number of connections to request GET/SET. When scaling to more connections, TSoR constantly achieves lower latency than others. While scaling to more connections, queueing delay within the Redis application shows up, which takes the performance bottleneck from the network to the application internally and that is beyond the scope of this project. Thus, to evaluate network performance, we do not scale the number of multi-connections to a larger number.  

Figure \ref{fig::redis_single_rps} presents the throughput comparisons with the metric of Request Per Second (RPS) for SET, GET, and INCR operations. TSoR achieves higher than other solutions for all three operations. For SET operation, TSoR achieves about 42K RPS, 2.3x times higher than host-mode runC. 

Figure \ref{fig::redis_multi_get_rps} and \ref{fig::redis_multi_set_rps} show the throughput comparisons while increasing the number of connections. For both GET and SET operations, TSoR almost linearly scales while varying the number of connections from 1 to 3. For GET operation, TSoR achieves 42K RPS for a single connection and 108K RPS for three connections. While scaling to more connections, performance is bottlenecked by the queue buildup within Redis rather than the network.

\subsubsection{Node.js}
Node.js is a software platform for server-side networking applications which can act as a web server with support of HTTP and Socket. We use node.js image to set up a web server and then use Apache HTTP server benchmarking tool (ab) \cite{ab} to generate requests from the client side. The server and client run on different host machines.

Figure \ref{fig::nodejs_lat} shows the response latency while varying the document length returned by the Node.js web server. TSoR completes the request faster than others, especially when transferring documents with large sizes. For transferring a $64KB$ document, TSoR achieves 44\% lower latency than host-mode runC. Because RDMA-based data path provides  higher throughput with lower stack overhead. 

Figure \ref{fig::nodejs_rps} shows the throughput for transferring documents with different sizes. For a $64KB$ document, TSoR achieves 3.11x times higher throughput than host-mode runC, which is the performance upper bound for other solutions.

\subsubsection{Etcd}
Etcd is a distributed key-value store that uses the Raft to achieve strong consensus. We use the etcd image (v3.0.0) to set up a server and run the official benchmarking tool with  ten connections. The server and client run over the different host machines. We select five common test cases for etcd: \textit{lease} is the primitive for consuming client keep-alive messages.
 \textit{range} request is to get multiple keys. \textit{txn-put} is to write a single key within a transaction (txn). \textit{stm} is the implementation of software transaction memory. A \textit{watch-get} request tells etcd to notify the requester of getting to any provided keys.

Figure \ref{fig::etcd} shows the throughput as Request Per Second (RPS) while executing different benchmarking tests.  Due to the space limit, we only show comparisons with host-mode runC and kata in bridge network mode, which achieve higher performance than other baselines 
without loss of generality. To unify the scale of different test cases, we normalize the RPS of other baselines to TSoR's performance numbers. For five common test cases, TSoR achieves higher RPS than other baselines. For example, for a typical test case of \textit{txn-put}, TSoR's RPS is 1.49x times higher than host-mode runC. 




%% file: future_work.tex
\section{Discussion}


\noindent \textbf{UDP socket over RDMA.} Although this work focus on the TCP Socket over RDMA, we also implement the stack for UDP Socket over RDMA as UDP is used in some scenarios such as DNS resolution even though it is not used as frequently as TCP. Compared with the TCP socket, one difference is that we use an Unreliable Datagram (UD) Queue Pair (QP) to transfer UDP data. A UD QP can transmit and receive data to/from any other UD QP in a connection-less manner. So each node only creates one UD QP for all UDP traffic. 

\noindent \textbf{Inter-pod and Intra-pod communication on the same host}. In one host node, there are both inter-Pod and intra-Pod TCP communications. TSoR currently supports inter-Pod communication based on RDMA memory copy. Hardware offloading stack of RDMA can bring benefits to inter-Pod communication of one host as well as inter-host. For the Intra-pod communication, TSoR supports TCP socket kernel buffer Zero Copy by re-using the TCP client side socket's write ring buffer (data buffer in Section \ref{sec::shm-comm}) as TCP server side socket's read ring buffer and vice versa. So the server side socket could read directly from the client side write buffer without memory copy for client/server socket data exchange. 

\noindent \textbf{Offloading TSoR Service to SmartNIC}. Emerging SmartNICs encloses computing resources (e.g., a multicore processor and onboard DRAM) that hold the potential to offload TSoR service from the host CPU \cite{bluefield,liquid,schuh2021xenic,liu2019offloading}. For the current design, TSoR runs the TSoR service over the host CPU, which costs one CPU core in practice. We are working on offloading the RDMA service to SmartNIC, and then the host CPU core resource can be saved.     

%% file: conclusion.tex
\section{Conclusion}
TSoR is a high-performance RDMA-based container network solution. TSoR has an RDMA-based data plane and integrates with the Kubernetes control plane. 
When enabling TSoR in the Kubernetes cluster, TCP socket-based applications  transparently gain better network performance without any code changes. 
Our evaluation shows that TSoR achieves higher throughput and lower latency than other popular container networking solutions. 